\documentclass[aps,prapplied,reprint,groupedaddress]{revtex4-2}

\usepackage{graphicx}
\usepackage{float}
\usepackage{color}
\usepackage{amsmath} 
\usepackage[bookmarks=false]{hyperref}
\usepackage[dvipsnames]{xcolor}
\hypersetup{pdfborder={0 0 0},
    colorlinks=true,
    linkcolor=blue,
    citecolor=blue,
    urlcolor=blue}

\begin{document}

%\preprint{}

\title{Unidirectional spin Hall magnetoresistance and spin-orbit torques in HM$_1$/Co/HM$_2$ trilayer systems}


\author{Anastasiia Moskaltsova}
%\email[]{}
%\thanks{}
\affiliation{Center for Spinelectronic Materials and Devices, Department of Physics, Bielefeld University, Universit\"atsstrasse 25, 33615 Bielefeld, Germany}

\author{Denis Dyck}
%\email[]{}
%\thanks{}
\affiliation{Center for Spinelectronic Materials and Devices, Department of Physics, Bielefeld University, Universit\"atsstrasse 25, 33615 Bielefeld, Germany}

%\author{Jan Biedinger}
%%\email[]{}
%%\homepage[]{Your web page}
%%\thanks{}
%%\altaffiliation{}
%\affiliation{Center for Spinelectronic Materials and Devices, Department of Physics, Bielefeld University, 33615 Bielefeld, Germany}

\author{Jan-Michael Schmalhorst}
%\email[]{}
%\thanks{}
\affiliation{Center for Spinelectronic Materials and Devices, Department of Physics, Bielefeld University, Universit\"atsstrasse 25, 33615 Bielefeld, Germany}

\author{G\"unter Reiss}
%\thanks{}
\affiliation{Center for Spinelectronic Materials and Devices, Department of Physics, Bielefeld University, Universit\"atsstrasse 25, 33615 Bielefeld, Germany}

\author{Timo Kuschel}
\email[]{tkuschel@physik.uni-bielefeld.de}
%\thanks{}
\affiliation{Center for Spinelectronic Materials and Devices, Department of Physics, Bielefeld University, Universit\"atsstrasse 25, 33615 Bielefeld, Germany}


\date{\today}

\begin{abstract}
We present a detailed analysis of harmonic longitudinal and Hall voltage measurements for in-plane magnetized Pt/Co/Ta and Ta/Co/Pt trilayers in reference to Pt/Co and Ta/Co bilayers. Enhancement of spin-orbit torques (SOTs) and unidirectional spin Hall magnetoresistance (USMR) is achieved by introducing the second heavy metal (HM) with the opposite sign of the spin Hall angle. The extracted SOT efficiencies are larger for the trilayers as compared to the bilayers, confirming the enhanced values reported for the trilayers with perpendicularly magnetized Co. The maximum effective spin Hall angle found for the Pt/Co/Ta trilayer reaches $\theta_{SH}$ = 20 \%. The USMR of the trilayer yields up to 27\% higher effect as for the respective bilayers with the largest effective USMR amplitude of -0.32 $\times$ 10$^{-5}$ for the Pt/Co/Ta trilayer at a charge current density of 10$^{7}$ A/cm$^2$. 
\end{abstract}

%\keywords{}

\maketitle

\section{Introduction}
%\subsection{}
%\subsubsection{}
 
Magnetoresistive (MR) effects play an important role in today's spintronic research \cite{Sander2017,Zheng2019}. Anisotropic magnetoresistance, giant magnetoresistance and tunneling magnetoresistance are widely used for memory applications \cite{Ikeda2007,Cai_2017} or as sensing devices \cite{Reig_2009}.

Recently, a new MR effect has been found in heavy metal (HM)/ferromagnetic insulator (FMI) hybrid structures called the spin Hall magnetoresistance (SMR). It has been intensively studied \cite{Nakayama,Vilestra,Chen,Althammer} with additional discussions and reports on HM/ferromagnetic metal (FMM) hybrids \cite{Kim2016, Kobs, Karwacki}. The SMR is based on the spin Hall effect (SHE) \cite{Dyakonov,Hirsch,Kato,Hoffmann}, which occurs in the HM and converts a charge current into a perpendicular spin current. Due to the inverse SHE, part of the spin current is converted back to the charge current, after it has been reflected from the HM/FMM(FMI) interface. The SHE emerges as a result of the strong spin-orbit coupling in HM materials such as Ta, Pt, W etc.

The perpendicular spin current can cross the interface from HM to FMM, but also from HM to FMI, thus converting from a spin-polarized charge current to a magnon spin current. Depending on the FM magnetization direction with respect to the spin polarization direction at the HM interface, the spin current can produce a torque on the FM's magnetization. If this SHE-induced spin-orbit torque (SOT) is strong enough, it can even switch the FM magnetization \cite{Miron,Liu}. The SOT magnetization switching is at present intensively studied due to its potential use for memory and logic applications \cite{Manchon2019,Chi2021,Sun2021}.

Nowadays, primarily spin-transfer torque (STT) \cite{Ralph} magnetization switching is used in magnetic tunnel junctions. However, the comparison between the SOT and STT- based MRAM is not straightforward, as was discussed by Grimaldi \textit{et al.} \cite{Grimaldi}. The SOT-MRAM requires an improvement of the critical current density to achieve high effectiveness, comparable to STT. However, the combination of both SOT and STT switching, so-called SHE-assisted STT switching can be advantageous and a key to low-power, ultra-fast and high density memory applications \cite{Cai_2017, Ahmed, Cai2021}. Recent reports discussed that the insertion of light elements or oxidization can enhance the SOT efficiency \cite{Qiu2015,Hibino2017,Hasegawa2018,Feng2020,
Feng2021,Anadon2021}. One reason could be that additional orbital Hall effects can occur and support the SOT switching \cite{Ding2020}.

Some of the MR effects depend quadratically on the magnetization \textbf{m}, although the physical origin differs from one MR effect to the other. However, Avci \textit{et al.} explored a new MR effect in Ta/Co/AlO$_{x}$ and Pt/Co/AlO$_{x}$ called unidirectional spin Hall magnetoresistance (USMR) that is linear in \textbf{m} and linear in the charge current density \textbf{j} \cite{AvciNature}. Up to now, several theoretical studies have been conducted to determine the USMR origins in various metallic bilayer systems. The first reports on HM/FMM thin films associate the USMR with the spin-diffusion theory \cite{Zhang}, in which the linear dependence on  \textit{\textbf{j}} is due to the SHE-induced spin accumulation at the HM/FMM interface. This spin accumulation can be described as $R^{2\omega}_{USMR}$ $\sim$ $\vert$\textit{\textbf{j}} $\times$ \textit{\textbf{m}}$\vert$, with $R^{2\omega}_{USMR}$ being the second harmonic longitudinal resistance term corresponding to USMR. The recent works on similar systems, e.g. permalloy/Pt bilayers \cite{Borisenko2018}, discuss that USMR is likely dominated by the dipolar magnons, whereas Avci \textit{et al.} broaden the understanding of the origins of the effect, attributing it to three different mechanisms: bulk and interface spin-dependent USMR as well as spin-flip USMR \cite{Avci2018}. For the case of HM/FMI bilayers, Sterk \textit{et al.} report USMR due to pure magnonic contributions \cite{Sterk2019}.

The USMR makes it possible to determine the in-plane direction of the magnetization or applied current and has a potential use as a magnetization read out mechanism in a multi-state memomry device \cite{AvciAPL110}. 
After the first report of USMR in HM/FMM bilayers, there have been several follow-up studies including large USMR in FM/paramagnetic GaMnAs bilayers \cite{Olejnik} and in topological insulator/FM heterostructures \cite{LV}, as well as in Heusler compound/HM bilayers \cite{Lidig2019}. A comprehensive investigation of the thickness dependence of USMR in Pt/Co bilayers has been performed by Yin \textit{et al.} \cite{Yin2017}. Additionally, Hasegawa \textit{et al.} reported enhancement of the USMR in Pt/Co system with a Cu interlayer \cite{Hasegawa2021}.

The magnetic proximity effect (MPE) can potentially influence the SOT efficiencies and magnetoresistance effects, as was shown for various HM/FMM bilayer systems \cite{Zhang2015, Peterson2018, Zhu2018}. Although, some studies claim irrelevance of the MPE in the bilayer systems with thin Co layer \cite{Zhu2018} it is essential to know the strength of the MPE for a clear picture of the magnetic properties \cite{Moskaltsova2020}.

In this work, we present a detailed study of the bi- and trilayer structures of HM/FM and HM$_1$/FM/HM$_2$, respectively. By switching the order of the HM$_1$ and HM$_2$, we can manipulate the HM/FM interfaces and thus tune both USMR and SOTs. As it was predicted by Zhang and Vignale \cite{Zhang}, one expects the enhancement of the USMR in such heterostructures with the FM sandwiched between two HMs with the opposite sign of the spin Hall angle $\theta_{SH}$. In such a case, an enhancement of the USMR is expected since both HMs generate spin currents with the same spin orientation, which can be controlled by the charge current direction and/or sign of the $\theta_{SH}$. It has been shown already for out-of-plane magnetized Co in a Pt/Co/Ta trilayer that one can tailor the stack and HM/FM interfaces to achieve higher effective $\theta_{SH}$ \cite{Woo}. Here, we investigate this approach for in-plane magnetized trilayers.

The sample stacks studied in this work are: sub/Pt 6/ Co 2.5/MgO 2/TaO$_\text{x}$ and sub/Ta 6/Co 2.5/MgO 2/TaO$_\text{x}$, further mentioned as Pt/Co and Ta/Co bilayers, respectively. The trilayers consisted of sub/Pt 6/ Co 2.5/Ta 6/MgO 2/TaO$_\text{x}$ and sub/Ta 6/Co 2.5/Pt 6/MgO 2/TaO$_\text{x}$, mentioned here as Pt/Co/Ta and Ta/Co/Pt, respectively. The thicknesses of the layers are indicated in nm. The substrate (sub) used for all the samples was Si/SiO$_{2}$ with an SiO$_{2}$-thickness of 50\,$\mu$m. The proximity induced moments have been already studied and are published in Ref. \cite{Moskaltsova2020}. The extracted thicknesses of the Pt spin-polarized layer for the trilayer samples are 0.9\,nm and 0.8\,nm for Ta/Co/Pt and Pt/Co/Ta, respectively. The induced Pt magnetic moments of 0.56\,$\mu_B$ per Pt atom at the interface for Ta/Co/Pt and 0.42\,$\mu_B$ per Pt atom at the interface for Pt/Co/Ta have been extracted \cite{Moskaltsova2020}. 
\\

\section{Experimental details}
\begin{figure}[h]
\includegraphics[width=\columnwidth]{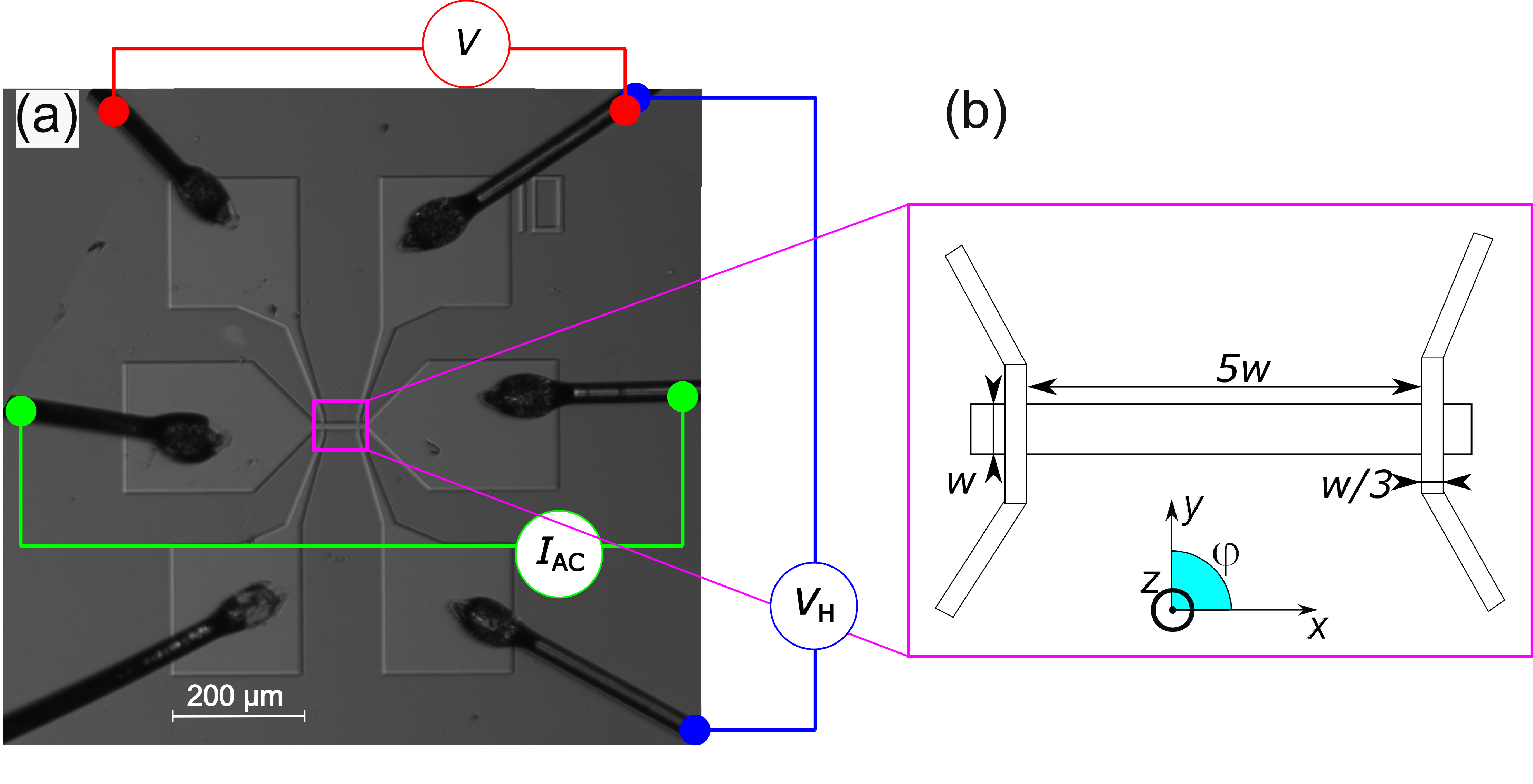}%
\caption{(a) Microscope image of a 10\,$\mu$m wire-bonded Hall bar and electrical connections. (b) Schematic representation of the zoomed-in region with the geometry definitions.  \label{Hbar}}
\end{figure}

The samples were prepared by DC magnetron sputtering (except for MgO, for which RF sputtering was used) on thermally oxidized Si substrates at room temperature. The top MgO/TaO$_\text{x}$ is a capping layer to protect the stacks from oxidation. The topmost TaO$_\text{x}$ layer was obtained by natural oxidation of a 2\,nm metallic Ta film. In-plane and out-of-plane saturation fields, as well as saturation magnetizations, were determined from vibrating sample magnetometry (VSM) loops measured up to 4\,T at room temperature (see Ref. \cite{Moskaltsova2020}). The samples are in-plane magnetized with $M_{s}$ = 1500\,kA/m for Pt/Co and 1400\,kA/m for all the other samples. The in-plane saturation fields are 0.4\,T for both trilayers and the Ta/Co bilayer, while Pt/Co exhibits $B_s=1.0\,\textrm{T}$. The coercive fields are in the range between 6.8 and 13\,mT. The films were further patterned via one-step electron beam lithography into Hall bar structures with a width $w$ and lengths $l=5\,w$, as shown in Fig. \ref{Hbar}.

The harmonic longitudinal and transverse voltage measurements are used here to characterize the spin dependent transport in the samples. The rotational measurements in \textit{xy} plane were carried out inside a dual Halbach cylinder array with a rotating magnetic field in the range from 0.05 to 1\,T (MultiMag, Magnetic Solutions Ltd.). By using two Signal Recovery SR7230 Lock-In amplifiers with two demodulators, we were able to measure simultaneously both first ($\omega$) and second (2$\omega$) harmonics of longitudinal and transverse (Hall) voltages. The measurements were performed at room temperature at a modulation frequency of \textit{f} = ${\omega/2\pi}$ = 1176\,Hz and with varied AC current amplitudes up to 2.5\,mA. Anomalous Hall resistances $R_{AHE}$ are obtained from magnetic field loops measured up to 4\,T at room temperature. \\

\section{Results and discussion} 
\subsection{Harmonic Hall voltage analysis: SOTs and thermal effects}

\begin{figure}[h]
\includegraphics[width=\columnwidth]{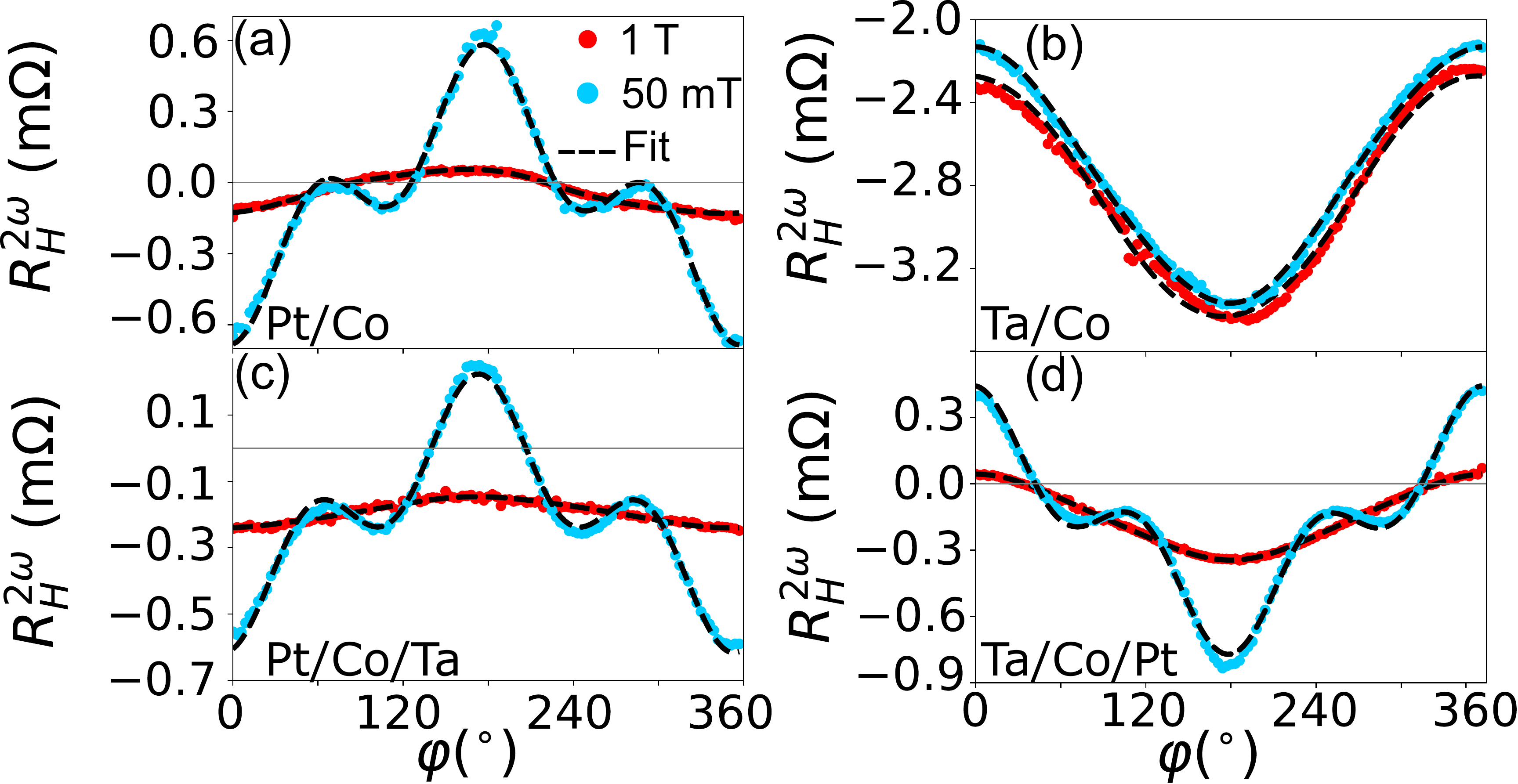}%
\caption{Second harmonic Hall resistances for the studied (a), (b) bi- and (c), (d) trilayers with a Hall bar width of $w=4\,\mu \textrm{m}$. Red and light blue circles correspond to the measured data at 1 and 0.05\,T, respectively. The fits (black dashed lines) are based on Eq. (\ref{eq1}). \label{Fig3}}
\end{figure}

First, we present the raw data and fits for the second harmonic transverse (Hall) resistance $R_{H}^{2\omega}$ measured as a function of the in-plane rotation angle $\varphi$ with respect to the external magnetic field. Figure \ref{Fig3} shows the data for the bi- and trilayers. The data analysis was based on the method developed by Avci \textit{et al.} \cite{AvciNature,AvciPRB}.

Following the analysis method, we first focus on the transverse second harmonic voltage, to obtain the information on the SOTs as well as to extract the transverse thermal effects. These effects arise from spin Seebeck and anomalous Nernst effects (SSE and ANE respectively) and must be taken into account. The second harmonic Hall voltage can be described as  \cite{Lau,Lukas}

%\begin{equation}
%\begin{dmath}
\begin{multline}
V^{2\omega}_{H}=[-\frac{B_{FL+Oe}}{B_{ext}}\,R_{PHE}\,\cos 2\varphi + \\ (-\frac{1}{2}\frac{B_{DL}}{B_{total}}\,R_{AHE}+\alpha^{\prime} I_{0})]\,\textit{I}_{rms}\,\cos \varphi. \label{eq1}
\end{multline}
%\end{dmath}
%\end{equation}
$B_{FL+Oe}$ is the field-like (FL) SOT and Oersted field contribution, while $B_{DL}$ corresponds to the damping-like (DL) SOT field. $R_{PHE}$ and $R_{AHE}$ are the planar Hall effect (PHE) and the anomalous Hall effect (AHE) resistances, respectively. The term $B_{total}$ is the total magnetic field. The term $\alpha^{\prime}$ $I_{0}$ describes any thermal contribution arising from SSE/ANE and $I_{rms}$ is the root mean square amplitude of the applied AC charge current. Thus, we can describe the transverse second harmonic resistance as a sum of the contributions

\begin{equation}
R_{H}^{2\omega}=R_{DL}^{2\omega} + R_{\nabla T}^{2\omega} + R_{FL+Oe}^{2\omega}. \label{eq2}
\end{equation}
The term $R_{\nabla T}^{2\omega}$ corresponds to the thermal effects driven by out-of-plane thermal gradients (SSE, ANE). As one can see from Eq.\,(\ref{eq1}), the first  term has a ($2\,\text{cos}^{3}\,\varphi - \text{cos}\,\varphi$) dependence, while the remaining two terms follow a cos $\varphi$ angular dependence.

\begin{figure}[h]
\includegraphics[width=\columnwidth]{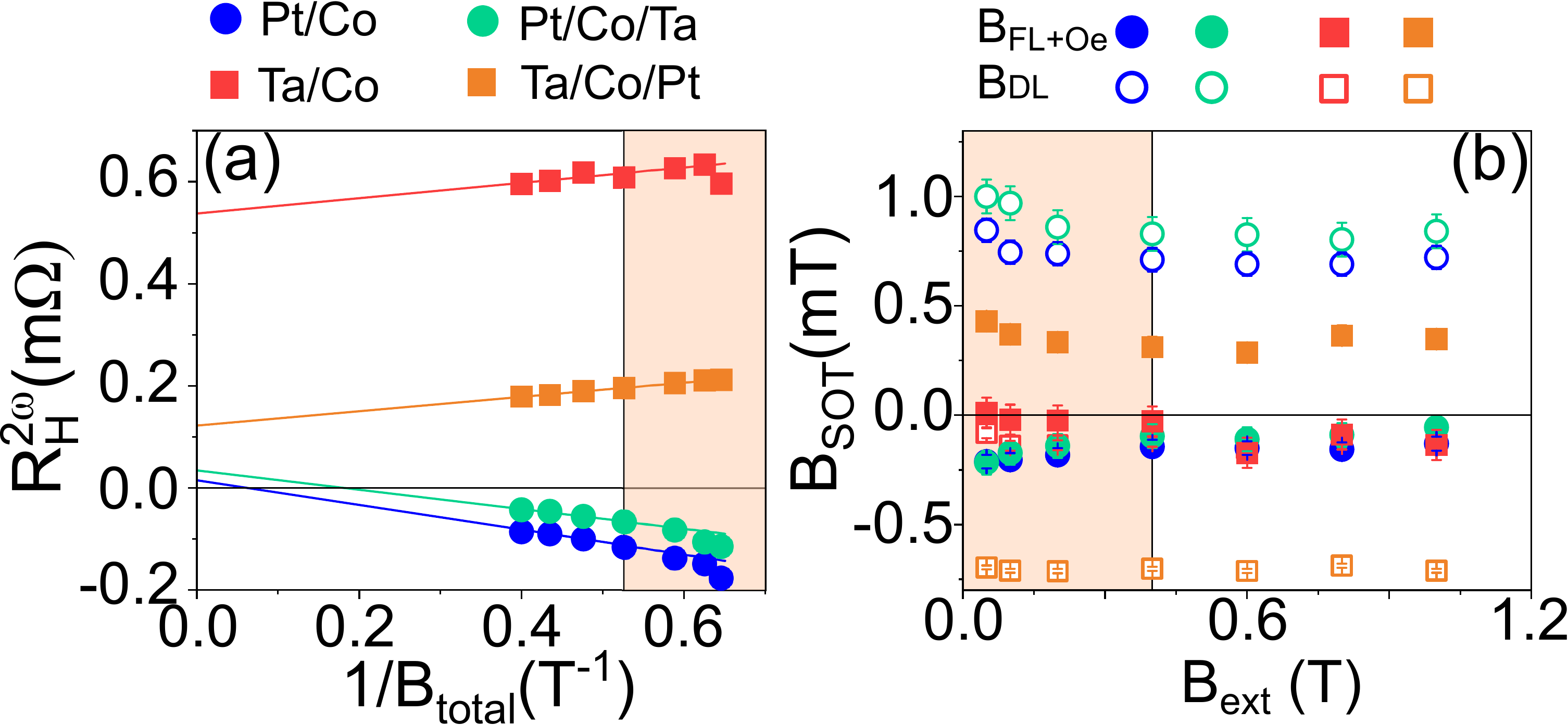}%
\caption{(a) $R_{H}^{2\omega}$ cos $\varphi$ contribution plotted against the inverse total magnetic field for the bi- and trilayers. Linear fits performed for the in-plane saturated regime $\textgreater$ 0.4\,T. (b) FL and DL SOTs as a function of the external magnetic field. The orange shaded area encloses the data in unsaturated regime.\label{Fig4}}
\end{figure}

We first fit the $R_{H}^{2\omega}$ raw data with a $a$\,cos\,$\varphi$ + $b$\,($2\,\text{cos}^{3}\,\varphi - \text{cos}\,\varphi$) function. We then extract the cos $\varphi$ contribution, which corresponds to the sum of DL SOT and thermal effects, as well as the (2 cos$^{3} \varphi$ - cos $\varphi$) contribution to extract FL SOT.
After the extraction of the cos $\varphi$ component, we can further separate the DL and thermal contributions by plotting the corresponding fit parameter against the inverse total magnetic field 1/$B_{total}$, with $B_{total}$= $B_{ext}$ + $B_{demag}$ - $B_{anis}$. The term ($B_{demag}$ - $B_{anis}$) can be identified as the out-of-plane saturation field. The example of such a separation is shown in Fig.~\ref{Fig4}(a) for the bi- and trilayers. The DL SOT is expected to have no influence in the large field regime. Therefore, the only contribution left at large magnetic fields is related to the thermal effects. The $B_{DL}$ is extracted from the slope of the linear fits depicted in Fig.~\ref{Fig4}(a). Thus, we define $R_{\nabla T}^{2\omega}$ as a \textit{y}-axis intercept of the 1/$B_{total}$ plotting. For further calculations of the USMR, we then convert the obtained $R_{\nabla T}^{2\omega}$ via the geometrical factor $l/w$ = 5 and subtract it from the raw longitudinal $R^{2\omega}$.

To calculate the DL and FL SOT fields ($B_{DL}$ and $B_{FL+Oe}$, respectively) we again utilize Eq. (\ref{eq1}). Here, the $B_{FL+Oe}$ term corresponds to a sum of FL SOT and Oersted field contributions. The Oersted field originates from the charge current flowing in the Pt and/or Ta layer(s) and can be estimated as $\mu_0 I_{rms}$/2$w$, with $\mu_0$ as the vacuum permeability, $I_{rms}$ as the root-mean-square current amplitude and $w$ as the width of the current line. $B_{Oe}$ was estimated to be -0.39\,mT and was subtracted from $B_{FL+Oe}$ for the FL SOT efficiency calculations. 
The SOT fields for bi- and trilayers are shown in Fig. \ref{Fig4}(b). As expected, there is no magnetic field dependence, except for the unsaturated regime in the magnetic field range $\textless$ 0.4\,T. We calculate the SOT efficiencies ($\xi_{DL}$ and $\xi_{FL}$) as \cite{Khvalkovskiy}

\begin{equation}
\xi_{DL(FL)}=\frac{2e}{\hbar}B_{DL(FL)}\frac{M_{s}t_{FM}}{j_{HM}}, \label{eq3}
\end{equation}

\noindent with \textit{e} being the electron charge, $\hbar$ the reduced Planck constant, $M_{s}$ and $t_{FM}$ the saturation magnetization and thickness of the FM, respectively. $j_{HM}$ is the charge current density amplitude in a HM layer.
 
Using the DL SOT values, we calculate the spin Hall angle $\theta_{SH}$, following a model in which the DL torque is entirely ascribed to the absorption of the spin current produced by the bulk spin Hall effect in the HM \cite{Liu,AvciAPL107}. In the case of bilayers, $\xi_{DL}$ = $\theta_{SH}$. We thus obtained $\theta_{SH}^{Pt}$ = 10\%, $\theta_{SH}^{Ta}$ = -3.5\% which is in agreement with the literature \cite{AvciPRB, Sagasta}. For the trilayers, however, we are only able to extract the effective spin Hall angle $\theta^{eff}_{SH}$, to which both Pt and Ta contribute. The obtained values yield 20\% for Pt/Co/Ta and -13\% for Ta/Co/Pt (see Table \ref{Table1}). The lower value for the latter trilayer may be attributed to the Ta oxidation at the substrate interface or due to a lower interface quality. The $\theta^{eff}_{SH}$ values obtained for the trilayers in this work are slightly lower than those previously reported (34\%) for the same trilayer structures with perpendicularly magnetized Co \cite{Woo}.
\begin{table*}
\caption{A summary of the results for the Pt and Ta systems.}\label{Table1}
\setlength{\tabcolsep}{6pt}
\begin{tabular}{@{}|l|l|l|l|l|l|l|l|l|l|l|l|@{}}
   \toprule 
    sample & \(R_{AHE}\) & \(R_{PHE}\) & \(B_{DL}\) & \(B_{FL+Oe}\) & \(\xi_{DL}\) & \(\xi_{FL}\)& \(R_{USMR}^{2\omega}\) & \(\frac{R_{USMR}}{R_{HM}}\) \(\times\) 10\(^{-5}\)\\
           & ($\Omega$) & ($\Omega$) & {(mT)} & {(mT)} & & & m$\Omega$& per 10$^{7}$ A/cm$^{2}$\\
    \hline
    Pt/Co & 0.35& 0.11 & 0.721 & -0.194 & $\sim$0.1 & 0.0006& -0.79 &-0.26\\
	\hline
    Ta/Co & 1.1 & 0.12 & -0.133 & -0.136 & $\sim$-0.035 & 0.013&0.95 & 0.11\\
    \hline
    Pt/Co/Ta & 0.23& 0.07 & 0.84 & -0.056 & $\sim$0.2 & -0.011& -0.66 &-0.32\\
    \hline
    Ta/Co/Pt & 0.24& 0.05& -0.71 & 0.335 & $\sim$-0.13 & 0.07&0.23 & 0.14\\
\botrule
\end{tabular}
\end{table*}
\subsection{Harmonic longitudinal voltage analysis: USMR comparison in bi- and trilayers}

\begin{figure}[h]
\includegraphics[width=\columnwidth]{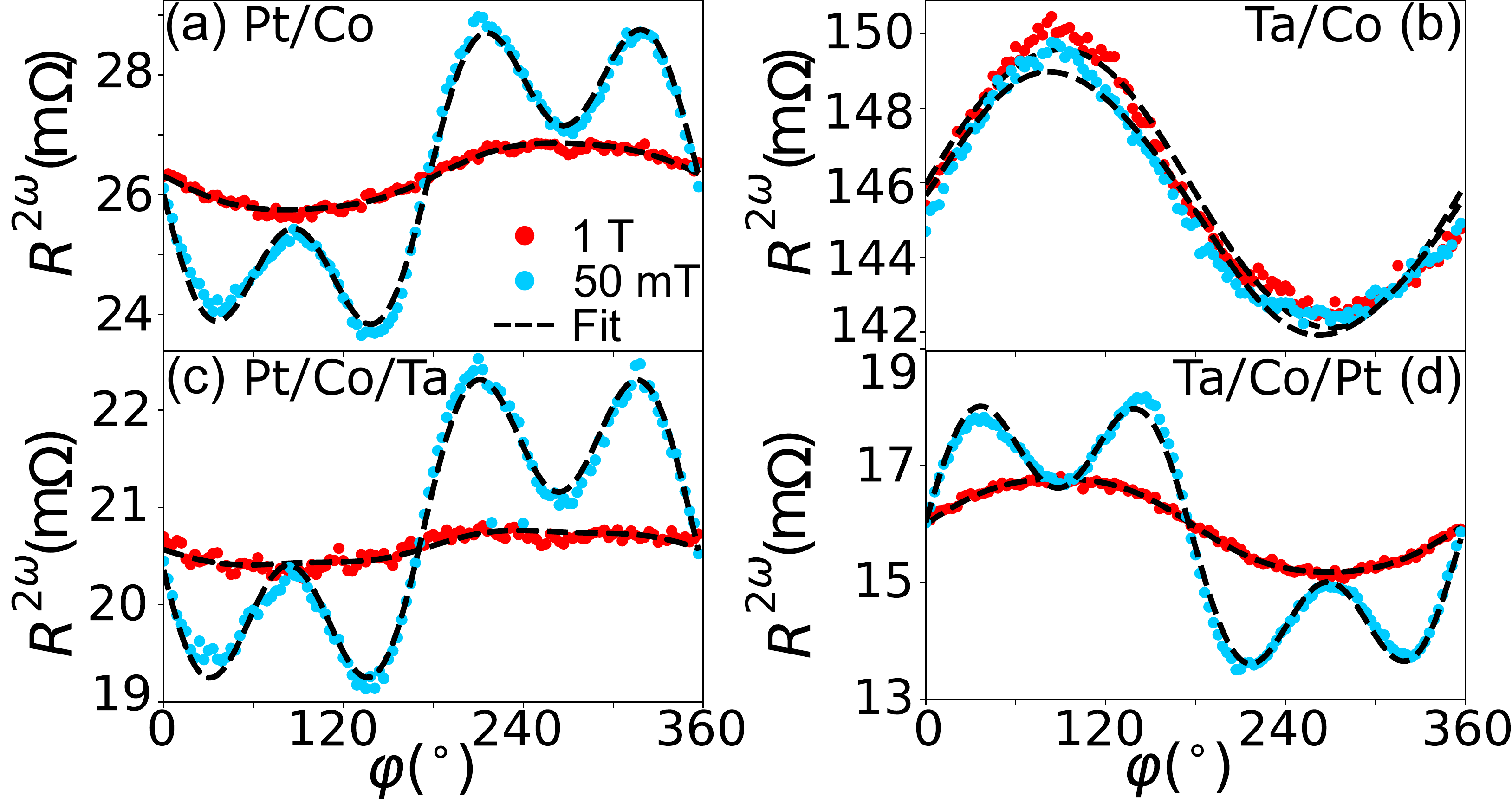}%
\caption{Longitudinal second harmonic resistance for the studied (a), (b) bi- and (c), (d) trilayers with a Hall bar width of $w$ = 4$\mu$m. Red and light blue circles correspond to the measured data at 1\,T and 0.05\,T, respectively. The fits (black dashed lines) were performed with the $a$\,sin $\varphi$ + $b$\,(2 cos $^{3} \varphi$ - cos $\varphi$) function. \label{Figp}}
\end{figure}

After the detailed analysis of the second harmonic Hall voltages, we take a closer look at the second harmonic longitudinal voltages. The corresponding data and fits are shown in Fig. \ref{Figp}. We observe a similar behavior as for the transverse case, in which for the samples containing Pt, extrema arise at lower magnetic fields at about $\varphi$ = 45$^{\circ}$, 135$^{\circ}$, 225$^{\circ}$, 315$^{\circ}$.

In order to investigate the influence in a different material with negative $\theta_{SH}$, but lower resistivity as compared to Ta, two samples with W instead of Ta were prepared. The W samples stacks are: Si/SiO$_{2}$/W 7/ Co 2.5/MgO 2/TaO$_\text{x}$ and Si/SiO$_{2}$/W 8/Co 2.5/Pt 6/MgO 2/TaO$_\text{x}$. These samples have been measured and analyzed in the same manner as the aforementioned Pt and Ta systems.

\begin{figure}[h]
\includegraphics[width=\columnwidth]{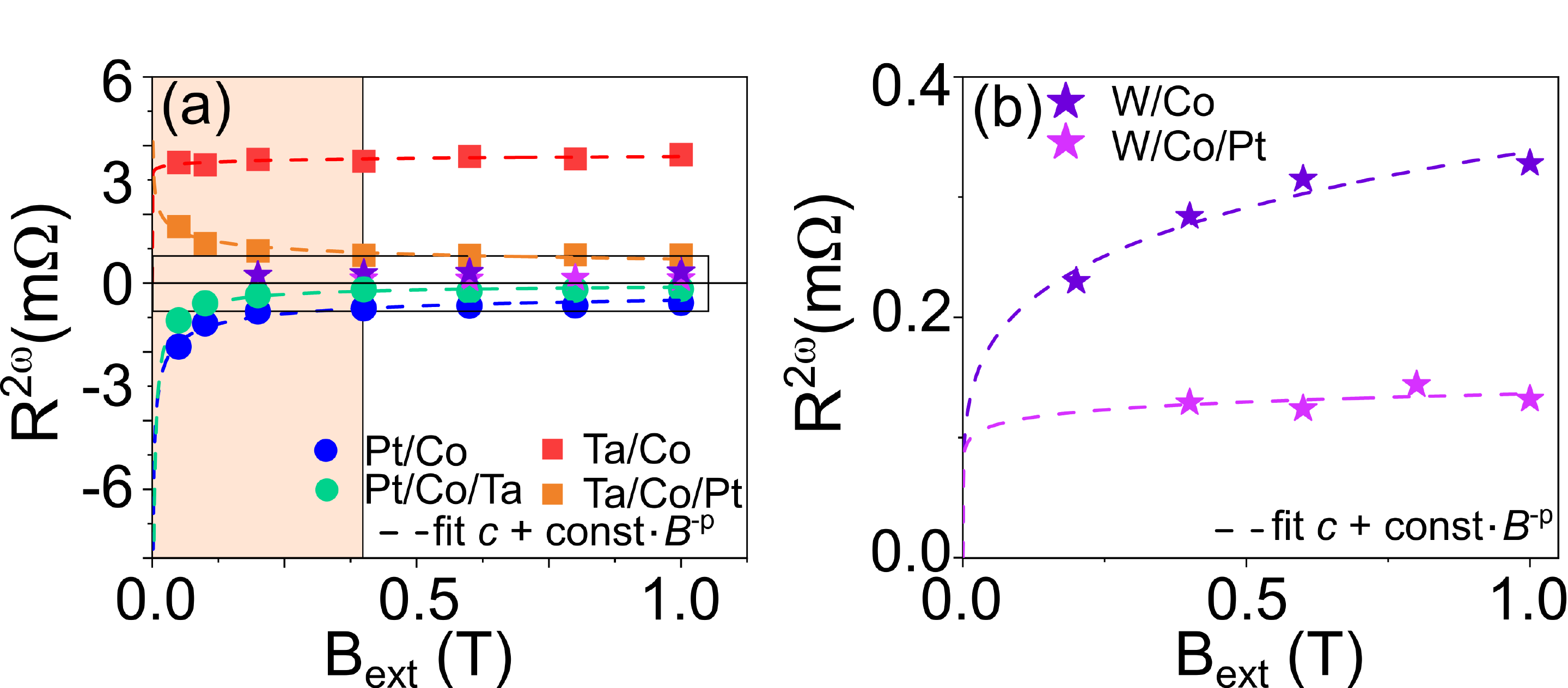}%
\caption{(a) Second harmonic longitudinal resistance dependence on magnetic field for bi- and trilayers with Pt, Ta and W. Dashed lines are fit with the $c$ + const $\cdot$ B$^{-p}$ function. The orange shaded area encloses the data in unsaturated regime. (b) Zoomed-in region with focus on W samples.\label{Fig5}}
\end{figure}

The $R^{2\omega}$ magnetic field dependence of the Pt and Ta samples is plotted in Fig. \ref{Fig5}(a). The samples containing W do not show a comparable $R^{2\omega}$ magnetic field dependence, as shown with magenta and violet stars in Fig. \ref{Fig5}(b). Following Ref. \cite{Avci2018}, we fit the data with a $c$ + const $\cdot$ B$^{-p}$ function (dashed line), with $c$ being an offset. It is clear, that in the saturated regime, above 0.4\,T, there is no dependence on the magnetic field. However, for lower magnetic field strengths (unsaturated state) the effect magnitude varies, although this behavior predominantly appears in Pt samples (blue and turquoise circles, as well as orange squares) with FL SOTs, hence, there is no magnetic field dependence for the Ta/Co bilayer (red squares). This finding confirms that the USMR effect is magnetic field-independent and the only reason for the higher signal amplitudes at lower fields is the FL SOT contribution.
%%%

\begin{figure}[h]
\includegraphics[width=\columnwidth]{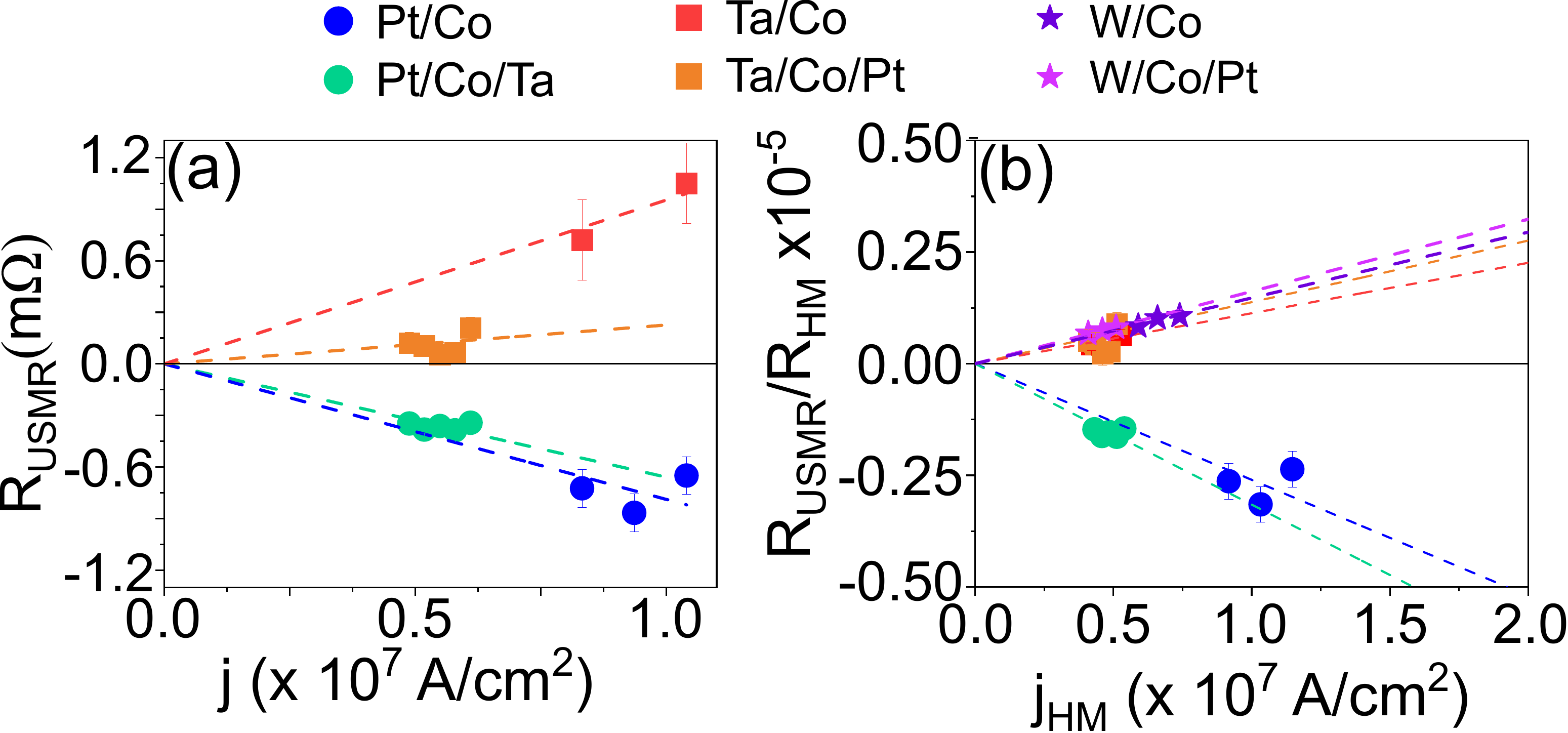}%
\caption{(a) USMR dependence on total charge current for bi- and trilayers with Pt and Ta. (b) Normalized USMR dependence on HM charge current density normalized to the HM resistance. Dashed lines are linear fits with fixed zero offset.\label{Fig6}}
\end{figure}

Finally, the USMR is calculated via $R_{USMR}^{2\omega}$ = $R^{2\omega} - \frac{l}{w} R_{\nabla T}^{2\omega}$, with $\frac{l}{w}$ = 5. A detailed analysis of the $R_{USMR}^{2\omega}$ current dependence is shown in Fig. \ref{Fig6}(a). For further comparison, the values at 1 $\times$ 10$^7$ A/cm$^2$ (slope of the linear fit) are discussed. It can be seen that for the Ta-based systems with positive USMR, the bilayer exhibits the highest value of 0.95\,m$\Omega$, while the respective trilayer yields $R_{USMR}^{2\omega}$ = 0.23\,m$\Omega$. In the case of the Pt based samples with negative USMR, the difference between the extracted USMR values is not large ($R_{USMR}^{2\omega}$ = -0.79\,m$\Omega$ vs. $R_{USMR}^{2\omega}$ =-0.66\,m$\Omega$), however, here, no enhancement for the trilayer is found. First of all, when we explore in more detail the Ta/Co bilayer, Ta in our film stacks has a resistivity 180 $\mu\Omega$\,cm, which is $\sim$\,4-5 times higher than Co (40 $\mu\Omega$\,cm) and, therefore, it is anticipated that most of the charge current flows through the Co layer. The resistivity of Co was taken from the literature for the same thin film thickness \citep{AvciPRB, Siniscalchi2020}, while the resistivities of Ta and Pt were calculated using the parallel resistor model with known Co resistivity. Thus, when Pt is added on top, the charge current distribution changes.

Thus, to correctly compare the USMR effect between the different samples, we take the ratio between the USMR effect $R_{USMR}^{2\omega}$ and the HM resistance $R_{HM}$. In case of the trilayers, with both Pt and Ta present, we calculate the current density flowing through both heavy metals using the parallel resistors model. Figure \ref{Fig6}(b) presents the current density dependence of the normalized USMR. We observe an enhancement of the effect in trilayers, of up to 27\%. The maximum effect of -0.32 $\times$ 10$^{-5}$ per 10$^7$ A/cm$^2$ is observed in the Pt/Co/Ta trilayer.

When we take a closer look at the bilayers, we see that the weighted USMR effect is stronger for Pt than Ta. Such a behavior can be explained by intermixing at the Ta/Co interface or oxidation of Ta at the substrate. By adding another HM interface to the bilayer, spin current is then driven from both sides into Co, and thus an increase of the USMR effect is achieved.

In case of the W-based samples, both bi- and trilayer show equal values within the uncertainty limit, as compared to the corresponding Ta samples. As was discussed by Pai \textit{et al.} \cite{Pai2012}, the $\theta_{SH}^W$ and resistivity strongly depend on the phase, with the highest $\theta_{SH}^W$ value being -33\% and resistivity in the range of 100-300\,$\mu\Omega$cm reported by the authors for the $\beta$ phase of tungsten. In this work, we determined the W resistivity to be $\approx$ 90\,$\mu\Omega$cm, indicating a mixture of both $\alpha$ and $\beta$ phases, thus a lower $\theta_{SH}^W$ is expected.

\section{Conclusions}
In summary, we have discussed harmonic voltage measurements of Pt/Co and Ta/Co bilayers, as well as their HM$_1$/Co/HM$_2$ trilayers. By analyzing the $2\omega$ Hall resistance, we were able to extract the important information about SOTs, as well as spin Hall angle for bilayers. Averaged extracted values of $\theta_{SH}^{Pt}$ $\approx$ 10\% and $\theta_{SH}^{Ta}$= -3.5\% are in agreement with previously reported ones. For the trilayers, effective $\theta_{SH}^{eff}$ are found to be $\theta_{SH}^{eff}$ = 20\% and -13\% for the Pt/Co/Ta and Ta/Co/Pt, respectively.
Finally, the USMR effect in the studied samples have similar values for bilayers, when compared to the literature. The obtained values for bilayers $R_{USMR}^{2\omega}/R_{HM}$ are -0.26 $\times$ 10$^{-5}$ and 0.11 $\times$ 10$^{-5}$ for Pt/Co and Ta/Co, respectively. For trilayers, we observe a modest enhancement, with $R_{USMR}^{2\omega}/R_{HM}$ found to be -0.32 $\times$ 10$^{-5}$ for Pt/Co/Ta and 0.14 $\times$ 10$^{-5}$ for Ta/Co/Pt samples per 10$^7$ A/cm$^2$, respectively. Additionally, two samples with W instead of Ta were measured, to compare the USMR effect size, as W is expected to have lower resistivity and higher negative spin Hall angle. The obtained USMR values for these samples, however, are equal to Ta-based systems within the uncertainty limit, thus no increase in USMR is observed. This is contrary to the expectations, but can be linked to the W-phase and film oxidation. These factors when combined can lead to the higher resistivity values, comparable to those of Ta. \\

%\appendix
%\section{}

\begin{acknowledgments}
The authors thank Daniel Carsten for help with initial experiments. We further thank Markus Meinert, Johannes Mendil and Pietro Gambardella for valuable discussions and comments. Finally, we thank Tristan Matalla-Wagner and Jan Biedinger for help with VSM and AGM measurements.\\
\end{acknowledgments}

\bibliography{references}

\end{document}